\documentclass[12pt, a4paper]{amsart}

\usepackage{cite}
\usepackage{color}
\usepackage[colorlinks,pdfusetitle,urlcolor=blue,citecolor=blue,linkcolor=red,bookmarksnumbered, plainpages=false]{hyperref}

\usepackage{amssymb,amsthm}
\usepackage{geometry}
\theoremstyle{definition}

\theoremstyle{remark}

\numberwithin{equation}{section}

\def\etal{\textit{et al\ }}

\begin{document}

\title[Dirac Particle with a Minimum Uncertainty in Position]{Free Motion of a Dirac Particle with a Minimum Uncertainty in Position}
\author{Arman Shokrollahi}
\address{Department of Mathematics, West Virginia University, Morgantown, WV, USA.}
\email{\href{mailto:arman@member.ams.org}{arman@member.ams.org}}


\begin{abstract}
In this paper, we present a covariant, relativistic noncommutative algebra which includes two small deformation parameters. Using this algebra, we obtain a generalized uncertainty principle which predicts a minimal observable length in measure of space-time distances. Then, we introduce a new representation for coordinate and momentum operators which leads to a generalized Dirac equation. The solutions of the generalized Dirac equation for a free particle will be explicitly obtained. We also obtain the modified fermionic propagator for a free Dirac particle.
\end{abstract}

\maketitle

\tableofcontents

\section{Introduction}

 The unification of quantum theory and general relativity is one of the most important tasks of the modern theoretical physics, although, there is not any unified, compatible theory yet. The first trouble to unify them is the loss of trajectory for particles in quantum theory and definition of path by geodesic equation in a Riemann curved space-time in general relativity. In fact, based on the Heisenberg electron microscope gedanken experiment, the process of determining the position of particles in quantum theory leads to an uncertainty in momentum of particles. So, there are some difficulties to formulate a quantum theory that includes gravity, because, a direct combination of quantum theory and general relativity gives a quantum gravity which is not a renormalizable quantum field theory.

  On the other hand, the modification of classical notion of the space-time is one of the common features of all quantum gravity theories. In these theories, it is assumed that the usual concept of continuity of the space-time manifold would break down when we probe distances smaller than the Planck length or energies larger than the Planck energy. If this conclusion can be confirmed by future experiments, it will make a remarkable influence on our understanding about our surrounding universe. It may help us to find the answer of many open problems such as the mechanism of singularity avoidance at early universe and also the black hole space-time.\\

 As a common feature of all quantum gravity candidates, the extrapolation of quantum theory and general relativity makes a minimal length in measurement of distances in the order of Planck length ($l_p=\sqrt{Gh/c^3}\sim 10^{-35}m$).   To have a simple intuition, if one considers the gravitational effects of scattered photon from an electron in a Heisenberg electron microscope, then an extra uncertainty is unavoidable in determination of the electron's position. This extra uncertainty is actually the source of finite resolution of space-time points. We note that this assumption is contacting the Heisenberg uncertainty principle which in principle agrees with the measurement of highly accurate results for particle positions or momenta, separately. In fact, in the Heisenberg picture, the minimum observable length is zero. So, if we are interested in incorporating the idea of minimal length, we need to modify the Heisenberg Uncertainty Principle to the Generalized Uncertainty Principle (GUP) \cite{acv,bu,bra1,dv1,gar,hoss1,kem1,kem2,mag1,mag2,mag3,noz1,noz3,noz7,ped,sca}. In other words, we should modify the commutation relations between position and momentum operators in the Hilbert space which leads us to a deformed Heisenberg algebra. In addition, the canonical commutation relations between position and momentum operators in Minkowski space can be modified to a curved space-time as
$$[x^\mu , p^\nu] = i\hbar g^{\mu \nu}(x),$$
 where $g^{\mu \nu}(x)$ is the measure (metric) of the curved space and includes the effects of the gravitational field. \\
 Moreover, String Theory,  which is one of the most promising scenarios that can unify all the fundamental interactions, guarantees a basic length in  order of $l_p$.  This theory appears as the main candidate for a quantum theory of the gravity and can logically solve some important problems such as ultraviolet divergence in a point-like quantum field theory of gravity; otherwise, the interaction of strings in the Planck energy scale and using an interpretation in terms of renormalization group leads us to a generalized uncertainty principle
 \begin{equation}
\Delta x \geq \frac{\hbar}{2 \Delta p}+\frac{\alpha}{c^3}G \Delta p.
\label{equ1}
\end{equation}

The relation above suggests a minimal length in measurement of distances. When the gravitational effects are unimportant, the Heisenberg uncertainty relation $\Delta x \geq \frac{\hbar}{2\Delta p}$ can be recovered. At the proximity of the Planck energy scale, the extra term in the generalized uncertainty principle (GUP) becomes relevant, and as has been indicated, we deal with a restriction of measurement in order of $l_p$ in resolution of distances and space-time points. This finite resolution of space-time points is due to this fact that a string cannot live in distances smaller than the string length. In fact, the existence of a minimum measurable length is one of the common properties of various candidates of quantum gravity such as string theory, loop quantum gravity and doubly special relativity. Also, some evidences from black hole physics assert that a minimal length of the order of the Planck length naturally arises from any theory of quantum gravity. In addition, in the sense of noncommutativity of the space-time manifold, we realize the existence of a minimal measurable length. \par
At the Planck energy scale,  the Schr\"odinger and Dirac equations, which are fundamental equations of non-relativistic and relativistic quantum theory, become converted to the generalized ones which contain the higher order space-time derivatives of the wave function. In fact, in GUP formalism, the idea of a minimum observable length and a maximum observable momentum changes the usual form of all Hamiltonians in quantum mechanics (see \cite{dv2} and references therein).  The modified Hamiltonians contain additional terms proportional to the powers greater than two of the momentum. So, in the quantum domain, the corresponding generalized Schr\"odinger or Dirac equation has a completely different differential structure.  As a consequence, some correctional terms will appear in energy spectrum and wave function relevant to the generalized equations.  There are some various works on incorporation of the quantum gravitational effects in the Schr\"odinger equation.  Kempf has expanded the studies on the structure of space-time at the Planck scale  \cite{kem1, kem3}. The modification of the Heisenberg algebra and its extension to higher dimensions
through GUP has been extensively studied recently (for more details see \cite{ben,bra1,bra2,kem2,lub,noz3,que3,ste}).  Hossenfelder et al. \cite{hoss1} obtained the first generalized Dirac equation without any further development.  Nozari \cite{noz5} found another generalization of Dirac equation in GUP framework and solved its eigenvalue problem for a free particle.  As a matter of fact, the deformed algebra, introduced in \cite{kem1,kem3}  and applied to quantum mechanical systems in \cite{ben,bra1,bra2,hoss1,kem2,lub,noz3,noz5,que3,ste}, is a non-relativistic one. Quesne and Tkachuk introduced a relativistic deformed algebra and determined its transformation properties under the (deformed) Poincar\'e algebra in \cite{que1, que2}. Their algebra is Lorentz covariant and contains Snyder algebra as a special case \cite{que1,sny}. \\

In this paper, we construct a \emph{relativistic} noncommutative algebra. This algebra contains two small deformation parameters and leads us to a generalized uncertainty principle predicting a minimal observable length in measurement of space-time distances. We introduce a new representation of coordinate and momentum operators to achieve a generalized Dirac equation. Then, we explicitly solve the generalized Dirac equation for a free particle. Finally, we calculate the modified fermionic propagator of a free single particle. \\

In section \ref{sec:Lorentz}, we review two types of non-relativistic and relativistic deformed algebras in quantized space-time and obtain the minimal uncertainties. Then, we define a new representation of contravariant $(D+1)$-vectors of coordinate and momentum using the Quesne-Tkachuk algebra. Applying this representation leads us to a generalization of Dirac equation in section \ref{sec:generalized}. We finally find a modified fermion propagator for our generalized Dirac equation in section \ref{sec:modified}, and the last section contains the conclusion.

\section{Lorentz covariant deformed algebra}
\label{sec:Lorentz}

For the first time, the deformed algebra leading to quantized space-time was introduced by Snyder in the relativistic case \cite{sny}. The deformed algebra suggested the existence of a finite lower bound for the possible resolution of length (minimal length). Kempf et al. \cite{kem1} obtained that minimal length in the $D$-dimension form from a deformed Heisenberg  algebra  whose  algebra  has been expressed as

 \begin{equation}
[X^i,P^j]=-i \hbar \left( g^{ij}(1+\beta P^2)-\beta^\prime P^i
P^j \right),
 \label{equ2}
 \end{equation}

  \begin{equation}
[X^i,X^j]=i\hbar \frac{2\beta - \beta^\prime
+(2\beta+\beta^\prime)\beta P^2}{1+\beta P^2} \left( P^i X^j - P^jX^i \right),
 \label{equ3}
 \end{equation}

  \begin{equation}
 [P^i,P^j]= 0,
 \label{equ4}
 \end{equation}
 where $X^i, P^j$  ($i,j\in\{1,2,\dots , D\}$) denote the generalized coordinate and momentum operators, respectively,  and $\beta , \beta^\prime$ are two very small nonnegative parameters of deformation. Note that the canonical commutation relations defined in Eqs. (\ref{equ2})-(\ref{equ4}) express a non-relativistic algebra. Quesne and Tkachuk modified the Kempf algebra to a relativistic one. Their algebra which is invariant under standard Lorentz transformation is expressed as

\begin{equation}
[X^\mu,P^\nu] = -i \hbar \left( g^{\mu\nu}(1-\beta P_\rho P^\rho) - \beta^\prime P^\mu P^\nu \right),
 \label{equ5}
 \end{equation}

 \begin{equation}
[X^\mu,X^\nu]= i\hbar \frac{2\beta-\beta^\prime
-(2\beta+\beta^\prime)\beta P_\rho P^\rho}{1-\beta P_\rho
P^\rho}(P^\mu X^\nu - P^\nu X^\mu),
 \label{equ6}
 \end{equation}

  \begin{equation}
[P^\mu,P^\nu]= 0,
 \label{equ7}
 \end{equation}
where $X^\mu,P^\nu$  ($\mu ,\nu \in\{0,1,2,\ldots , D\}$) are $(D+1)$-vectors and the parameters of deformation $\beta ,\beta^\prime$ considered very small.  In spite of the similarity of the commutative relations in Eqs. (\ref{equ5})-(\ref{equ7}) with Kempf's, the algebra proposed by Quesne and Tkachuk is a truly new one. Particularly, it has been reduced to the Snyder algebra (for more details see \cite{que1, que2}). Using this new algebra, we obtain the minimal observable length in measurement of space-time distances from a generalized uncertainty principle (GUP). This GUP is determined by choosing any pair of position and momentum components $X^i, P^i$  ($i \in\{1,2,\ldots ,D\}$) and applying these elements to Schwartz's inequality. After assuming isotropic uncertainty $\Delta P^j = \Delta P$, $j=1,2,\ldots ,D$,  the GUP becomes

 \begin{equation}
\Delta X^i \Delta P \geq \frac{\hbar}{2} \left| 1-\beta \left(
\langle (P^0)^2\rangle - 3(\Delta P)^2 - \sum_{j=1}^3 \langle P^j
\rangle ^2 \right)+ \beta^\prime \left( (\Delta P)^2 + \langle P^i
\rangle ^2 \right) \right|.
  \label{equ8}
 \end{equation}

 The corresponding minimal position uncertainty is

\begin{equation}
\Delta X^i_{\min}=\hbar \sqrt{(3\beta + \beta^\prime)\left\{ 1-\beta
\left(\langle(P^0)^2\rangle-\sum_{j=1}^3\langle P^j\rangle
^2\right)+\beta^\prime \langle P^i\rangle ^2 \right\}}.
 \label{equ9}
 \end{equation}

So, the smallest uncertainty in position has the value

\begin{equation}
(\Delta X^i)_0 = \hbar \sqrt{(3\beta +\beta^\prime)\left( 1-\beta
\langle (P^0)^2\rangle \right)}.
 \label{equ10}
\end{equation}

Now, we consider $\beta , \beta^\prime$ as small quantities of the first order. In this paper, we study only the case $\beta^\prime = 2\beta$ which leaves the commutation relations between the operators $X^\mu$ unchanged at the first order of $\beta$, i.e.,

\begin{equation}
[X^\mu , P^\nu] = -i \hbar \left( g^{\mu\nu}(1-\beta P_\rho P^\rho)
- 2\beta P^\mu P^\nu\right),
 \label{equ11}
\end{equation}

 \begin{equation}
 [X^\mu,X^\nu]= 0,
 \label{equ12}
 \end{equation}

  \begin{equation}
[P^\mu,P^\nu]= 0.
 \label{equ13}
 \end{equation}

The position and momentum operators can be represented by

 \begin{equation}
  X^\mu = x^\mu, \quad  P^\mu = (1-\beta p^2)p^\mu = i\hbar (1-\beta
  p^2)\partial ^\mu,
 \label{equ14}
 \end{equation}
 where $x^\mu , p^\mu$ are contravariant $(D+1)$-vectors in the conventional $(D+1)$-dimensional continuous space-time. The quasi-position representation showed in Eq. (\ref{equ14}) satisfies Eq. (\ref{equ11}) at the first order of $\beta$, and the simplest representation of $x^\mu , p^\mu$ is coordinate-diagonal
 \begin{equation}
 x^\mu = x^\mu, \quad  p^\mu = i\hbar \partial ^\mu.
 \label{equ15}
 \end{equation}

The commutation relations between $x^\mu , p^\nu$ can be displayed as

\begin{equation}
[x^\mu , p^\nu] = -i\hbar g^{\mu\nu}, \quad [x^\mu , p_\nu]=-i \hbar
\delta_{\nu}^\mu.
 \label{equ16}
\end{equation}


\section{Generalized Dirac wave equation}
\label{sec:generalized}

 The Dirac equation formulated by Dirac in $1928$ as an equation for the relativistic quantum wave function of a single electron, and interpreted as a field equation. The symmetrical form of the equation is

 \begin{equation}
 \left( \gamma^\mu p_\mu - mc \right) \Psi = 0,
 \label{equ17}
 \end{equation}
 where $\gamma^\mu$ are $4 \times 4$ matrices defined by
$$
 \gamma^0 = \hat{\beta} \ {\mbox{ and }} \ \gamma^i = \hat{\beta}\hat{\alpha}^i
 \qquad \ (i= 1,2,3)
$$
where if $I$ and $\mathcal{O}$ represent $2\times 2$ unit and block matrices, respectively, then
$$
 \hat{\beta } = \left(
                    \begin{array}{cc}
                      I & \mathcal{O} \\
                      \mathcal{O} & -I \\
                    \end{array}
                  \right), \quad
                  \hat{\alpha} = \left(
                    \begin{array}{cc}
                      I & \mathcal{O} \\
                      \mathcal{O} & -I \\
                    \end{array}
                  \right).
 $$

 Now, we want to study a generalized Dirac equation based on the Lorentz invariant deformed algebra introduced in Eq. (\ref{equ11}). Replacing $p^\mu$ in Eq. (\ref{equ17}) by $P^\mu$ defined in Eq. (\ref{equ14}) provides us with a Lorentz covariance extension of this equation. The generalized Dirac wave equation can be written as

 \begin{equation}
 \left( \gamma^\mu P_\mu - mc \right) \Psi = 0,
 \label{equ18}
 \end{equation}
or
 \begin{equation}
 \left[ i\hbar \gamma^\mu (1+\beta \hbar ^2 \partial^2)\partial_\mu -mc
 \right]\Psi = 0.
 \label{equ19}
 \end{equation}

Hence, the deformed Dirac equation depends on higher order derivatives of the wave function, and the conventional equation can be obtained when the deformation parameter vanishes. We examine the solution of the Dirac equation without potentials and rewrite the deformed equation in the following form

\begin{equation}
\left[ i\hbar (1+\beta \hbar^2
\partial^2)\frac{\partial}{\partial t} - c\boldsymbol{\alpha}\cdot \mathbf{p}(1+\beta \hbar^2
\partial^2) - mc^2 \hat{\beta}\right]\Psi=0.
 \label{equ20}
\end{equation}

Its stationary states are

\begin{equation}
 \Psi (\mathbf{r},t)=\Psi (\mathbf{r})\exp \left(-\frac{i}{\hbar}\epsilon
 t \right),
 \label{equ21}
 \end{equation}
where the quantity $\epsilon$ denotes the time evolution of the stationary state $\Psi (\mathbf{r})$. The four-component spinor splits up into two two-component spinors $\varphi$ and $\chi$, i.e.,
\begin{equation}
 \Psi = \left(
          \begin{array}{c}
            \Psi_1 \\
            \Psi_2  \\
            \Psi_3  \\
            \Psi_4  \\
          \end{array}
        \right) = \left(
                    \begin{array}{c}
                      \varphi \\
                      \chi \\
                    \end{array}
                  \right),
 \label{equ22}
 \end{equation}
where
\begin{equation}
\varphi = \left(
            \begin{array}{c}
              \Psi_1 \\
              \Psi_2 \\
            \end{array}
          \right), \qquad
          \chi = \left(
            \begin{array}{c}
              \Psi_3 \\
              \Psi_4 \\
            \end{array}
          \right).
 \label{equ23}
\end{equation}

 Using the explicit form of the $\hat{\alpha}$ and $\hat{\beta}$ matrices, Eq. (\ref{equ20}) can be expressed as
 \begin{equation}
 \left( 1-\beta \frac{\epsilon^2}{c^2} - \beta
 \hbar^2 \nabla^2\right) \epsilon \varphi = \left( 1-\beta \frac{\epsilon^2}{c^2} - \beta
 \hbar^2 \nabla^2\right) c \boldsymbol{\sigma} \cdot \mathbf{p} \chi + mc^2
 \varphi,
 \label{equ24}
 \end{equation}

 \begin{equation}
 \left( 1-\beta \frac{\epsilon^2}{c^2} - \beta
 \hbar^2 \nabla^2 \right) \epsilon \chi = \left( 1-\beta \frac{\epsilon^2}{c^2} - \beta
 \hbar^2 \nabla^2\right) c \boldsymbol{\sigma} \cdot
 \mathbf{p} \varphi + mc^2 \chi.
 \label{equ25}
 \end{equation}

States with definite momentum $\mathbf{p}$ can be written as
\begin{equation}
 \left(
   \begin{array}{c}
     \varphi \\
     \chi \\
   \end{array}
 \right) = \left(
   \begin{array}{c}
     \varphi_0 \\
     \chi_0 \\
   \end{array}
 \right)
 \exp \left( -\frac{i}{\hbar}\mathbf{p}\cdot
 \mathbf{r} \right).
 \label{equ26}
 \end{equation}

 Eqs. (\ref{equ24})-(\ref{equ25}) give the same equations for $\varphi_0$ and $\chi_0$, replacing the operators $\hat{p}$ and $\hat{p}^2$ by the eigenvalue $p$. So,  we have a linear homogenous system of equations for $\varphi_0$ and $\chi_0$, and it has nontrivial solutions only in the case of a vanishing determinant of the coefficients, that is

 \begin{equation}
 \det \left(
        \begin{array}{cc}
          \left( \epsilon \left( 1-\beta \frac {\epsilon^2}{c^2}+\beta\mathbf{p}^2\right) - mc^2 \right)I & -c
          \left(
          1-\beta \frac {\epsilon^2}{c^2}+\beta\mathbf{p}^2\right)(\boldsymbol{\sigma}\cdot \mathbf{p}) \\
          -c \left(
          1-\beta \frac {\epsilon^2}{c^2}+\beta\mathbf{p}^2\right)(\boldsymbol{\sigma}\cdot \mathbf{p}) &
          \left( \epsilon \left( 1-\beta \frac {\epsilon^2}{c^2}+\beta\mathbf{p}^2\right) + mc^2 \right)I\\
        \end{array}
      \right) = 0.
 \label{equ27}
 \end{equation}

Using the following expression

\begin{equation}
\left( \boldsymbol{\sigma} \cdot \mathbf{A} \right)
\left( \boldsymbol{\sigma} \cdot \mathbf{B} \right) =
\mathbf{A} \cdot \mathbf{B} + i
\boldsymbol{\sigma} \cdot \left( \mathbf{A} \times
\mathbf{B} \right),
 \label{equ28}
 \end{equation}
 Eq. (\ref{equ27}) can be written as

 \begin{equation}
 \left( \epsilon^2 - c^2 \mathbf{p}^2\right)\left( 1-\beta \frac{\epsilon^2}{c^2} + \beta
 \mathbf{p}^2\right)^2 - m^2 c^4 = 0.
 \label{equ29}
 \end{equation}

 Clearly when $\beta \rightarrow 0$, Eq. (\ref{equ29}) implies the following conventional result

 \begin{equation}
 \epsilon^2 = m^2c^4 + c^2 \mathbf{p}^2,
 \label{equ30}
 \end{equation}

 \begin{equation}
 \epsilon = \pm E_p,   \qquad   E_p = c \sqrt{\mathbf{p}^2 + m^2c^2}.
 \label{equ31}
 \end{equation}

 The two signs of the time-evolution factor $\epsilon$ in Eq. (\ref{equ31}) correspond to two types of positive and negative solutions of the Dirac equation. But for $\beta \neq 0$, neglecting terms of order $\beta^2$ provides us with the following two sets of results

\begin{equation}
\epsilon_{-} = \pm E_p^{(-)}, \qquad  E_p^{(-)} = c
\sqrt{\mathbf{p}^2 + \mu^2_{-}c^2},
 \label{equ32}
\end{equation}

 \begin{equation}
\epsilon_{+} = \pm E_p^{(+)}, \qquad  E_p^{(+)} = c
\sqrt{\mathbf{p}^2 + \mu^2_{+}c^2},
 \label{equ33}
\end{equation}
where $\mu_{-}$ and $\mu_{+}$ are two parameters with mass dimension as follows

\begin{eqnarray}
 \mu_{-}  = \frac{1}{2\sqrt{2\beta}c} \left\{ \sqrt{1 + 2 \sqrt{2\beta}mc}-\sqrt{1-\sqrt{2\beta}mc} \right\},
 \label{equ34}
\end{eqnarray}
\begin{eqnarray}
 \mu_{+}  = \frac{1}{2\sqrt{2\beta}c} \left\{ \sqrt{1 + 2 \sqrt{2\beta}mc}+\sqrt{1-\sqrt{2\beta}mc} \right\}.
 \label{equ3x}
\end{eqnarray}

To avoid having complex masses in the deformed field, we need to have

\begin{equation}
 2\sqrt{2\beta} mc < 1
 \label{equ35}
 \end{equation}
in Eqs. (\ref{equ34})-(\ref{equ3x}).  We have two masses in the generalized Dirac field. A quantum field theory with two kinds of mass can be found in \cite{gre}. From Eqs. (\ref{equ25})-(\ref{equ26}), for a fixed $\epsilon$,  we have that

\begin{equation}
 \chi_0 = \frac{c \left( 1-\beta\frac{\epsilon^2}{c^2} + \beta \mathbf{p}^2\right)
 (\boldsymbol{\sigma} \cdot \mathbf{p})}
 {\epsilon\left( 1-\beta\frac{\epsilon^2}{c^2} +
 \beta \mathbf{p}^2\right) + mc^2}\varphi_0,
 \label{equ36}
 \end{equation}
where the two-spinor $\varphi_0$ is
\begin{equation}
 \varphi_0 = U = \left(
                   \begin{array}{c}
                     U_1 \\
                     U_2 \\
                   \end{array}
                 \right),
 \label{equ37}
 \end{equation}
with the following normalization
\begin{equation}
 U^{\dag}U = U^{\ast}_1 U_1 + U^{\ast}_2 U_2 =1,
 \label{equ38}
 \end{equation}
and $U_1$, $U_2$ are complex. Using Eqs. ($\ref{equ21}$) and ($\ref{equ26}$), we obtain the complete set of positive and negative free solutions of the generalized Dirac equation as follows

\begin{equation}
 \Psi_{\mathbf{p}\lambda}^{(\mp)} (\mathbf{r},t) =
 \frac{N^{(\mp)}}{(2\pi\hbar)^{\frac{3}{2}}} \left(
                                               \begin{array}{c}
                                                 U \\
                                                 \frac{c\left( 1-\frac{\beta}{c^2} {E_p^{(\mp)}}^2 + \beta\mathbf{p}^2\right)
                                                 (\boldsymbol{\sigma} \cdot \mathbf{p})}
                                                 {\lambda E_p^{(\mp)}\left( 1-\frac{\beta}{c^2} {E_p^{(\mp)}}^2 + \beta\mathbf{p}^2\right)
                                                 + mc^2}U \\
                                               \end{array}
                                             \right)
      \exp \left({\frac{i}{\hbar} \left( \mathbf{p} \cdot \mathbf{r} - \lambda
      E_p^{(\mp)}t\right)} \right).
 \label{equ39}
 \end{equation}

Here $\lambda = \pm 1$ characterizes the positive and negative solutions with the time evolution factors $\epsilon_{(\mp)} = \lambda E_p^{(\mp)}$.  The normalization factors $N^{(\mp)}$ are determined from the following property

\begin{equation}
 \int \Psi_{\mathbf{p}\lambda}^{(\mp)^\dag}(\mathbf{r},t)
 \Psi_{\mathbf{p}\lambda}^{(\mp)} (\mathbf{r},t) d^3 r  =  \delta_{\lambda \lambda^{\prime}} \delta(\mathbf{p} -
 \mathbf{p}^{\prime}).
 \label{equ40}
\end{equation}

 By  Eq. (\ref{equ28}), the normalization factors can be obtained as

\begin{equation}
 N^{(\mp)} = \left\{ \frac{\left[ \lambda E_p^{(\mp)}\left( 1-\beta \mu^2_{(\mp)}c^2\right) + mc^2 \right]^2}{\left[ \lambda E_p^{(\mp)}
 \left( 1-\beta \mu^2_{(\mp)}c^2 \right) + mc^2 \right]^2 + c^2 \mathbf{p}^2
 \left( 1-\beta\mu_{(\mp)}^2 c^2\right)^2} \right\}^{1/2}.
 \label{equ41}
\end{equation}

Now we simplify $N^{(-)}$ in order to approach the conventional result. By keeping only the first order of $\beta$, the $\mu_{-}$ mass becomes
\begin{equation}
\mu_{-} \simeq m(1+\beta m^2 c^2).
 \label{equ42}
\end{equation}

Using Eq. (\ref{equ32}), we get the following expression
\begin{equation}
 E_p^{(-)^2} = c^2 \mathbf{p}^2 + m^2 c^4 + 2\beta m^4 c^6,
 \label{equ43}
\end{equation}
which is a modification of the well-known Einstein relation. For these results, we obtain the following generalized Dirac four-spinors

\begin{eqnarray}
 \Psi_{\mathbf{p}\lambda}^{(-)} (\mathbf{r},t)  &=&
 \frac{1}{(2\pi\hbar)^{\frac{3}{2}}} \sqrt{\frac{\lambda E_p^{(-)}(1 - \beta m^2 c^2) + m c^2}{2\lambda
E_p^{(-)}(1 - \beta m^2 c^2)}} \\
                                 & & \left(  \begin{array}{c}
                                                 U \\
                                                 \frac{c\left( 1-\beta m^2c^2\right)(\boldsymbol{\sigma} \cdot \mathbf{p})}
                                                 {\lambda E_p^{(-)}\left( 1 - \beta m^2 c^2 \right)
                                                 + mc^2}U \nonumber \\
                                               \end{array}
                                    \right)
      \exp \left({\frac{i}{\hbar} \left( \mathbf{p} \cdot \mathbf{r} - \lambda
      E_p^{(-)}t\right)}\right).
 \label{equ44}
 \end{eqnarray}

 It is clear that for $\beta =0$, the positive and negative free solutions of the conventional Dirac equation are obtained as follows  \cite{grein}

\begin{equation}
\Psi_{\mathbf{p}\lambda} (\mathbf{r},t) =
 \frac{1}{(2\pi\hbar)^{\frac{3}{2}}} \sqrt{\frac{\lambda E_p + m c^2}{2 \lambda E_p}}
                                      \left(
                                           \begin{array}{c}
                                                 U \\
                                                 \frac{c
                                                 (\boldsymbol{\sigma} \cdot
                                                 \mathbf{p})}
                                                 {\lambda E_p
                                                 + mc^2}U \\
                                               \end{array}
                                         \right)
      \exp \left(\frac{i}{\hbar} \left( \mathbf{p} \cdot \mathbf{r} - \lambda
      E_p t \right) \right).
\label{equ45}
\end{equation}

There is no similarity with the other set of the generalized solutions that contain $\mu_{+}$ in the usual Dirac field. So, in contrast to the ordinary answers, we have a modified set of negative and positive answers. In addition, there is a completely new set of positive and negative answers.

\section{Modified fermionic propagator}
\label{sec:modified}

The ordinary inhomogeneous Dirac equation proportional to Dirac delta function is

\begin{equation}
\left[ i\hbar \gamma^{\mu}\frac{\partial}{\partial x^{\mu}}-mc
\right]S_F (x) = \hbar \delta^4 (x),
 \label{equ46}
\end{equation}
where $S_F (x)$ is the fermion propagator, and the $3+1$ dimensions delta function is

\begin{equation}
 \delta^4 (x) = \frac{1}{(2\pi \hbar)^4} \int d^4 p \exp \left( -
 \frac{i}{\hbar}p \cdot x
 \right).
 \label{equ47}
\end{equation}

 The delta function $\delta (x)$ has dimension $L^{-4}$, and $\exp\left( -\frac{i}{\hbar}p \cdot x \right)$ is dimensionless. So the dimension of $S_F(x)$ is $L^{-3}$.  Suppose that
\begin{equation}
 S_F(x) = \frac{1}{(2\pi \hbar)^4} \int d^4 p \exp \left( - \frac{i}{\hbar}p \cdot
 x \widetilde{S}(p)
 \right),
 \label{equ48}
\end{equation}
where $\widetilde{S}(p)$ is the momentum representation of the fermion propagator. Substituting $S_F (x)$ in Eq. (\ref{equ46})  gives that

\begin{equation}
\widetilde{S}(p) = \hbar \frac{p+mc}{p^2 - m^2 c^2},
 \label{equ49}
\end{equation}
 and substituting Eq. (\ref{equ49}) in Eq. (\ref{equ48}) gives the following

\begin{equation}
S_F(x) = \frac{\hbar}{(2\pi \hbar)^4} \int d^4 p \exp \left( -
\frac{i}{\hbar}p \cdot x \right) \frac{p+mc}{p^2 - m^2 c^2}.
 \label{equ50}
\end{equation}

 The general form of the inhomogeneous Dirac equation is written as
 \begin{equation}
 \left[ i\hbar \gamma^{\mu}\frac{\partial}{\partial x^{\mu}} - mc \right] \Psi (x) = \hbar J (x),
 \label{equ51}
\end{equation}
 where $\Psi (x)$ and $J(x)$ are the spinor field and the external heterogeneity factor with dimensions $L^{-3/2}$ and $L^{-5/2}$, respectively. The general solution to Eq. (\ref{equ51}) can be expressed as
\begin{equation}
 \Psi (x) = \Psi_0 (x) + \int d^4 y S_F (x-y) J(y),
 \label{equ52}
\end{equation}
where $\Psi_0(x)$ is the answer of the homogeneous Dirac equation
\begin{equation}
\left[ i\hbar \gamma^{\mu} \frac{\partial}{\partial x^{\mu}} -
mc\right] \Psi_0 (x) = 0,
 \label{equ53}
 \end{equation}
 and
\begin{equation}
 S_F(x-y) = \frac{\hbar}{(2\pi \hbar)^4}\int d^4 p \exp \left( - \frac{i}{\hbar}p\cdot(x-y) \frac{p+mc}{p^2 - m^2c^2}\right),
 \label{equ54}
 \end{equation}
satisfies the following expression
\begin{equation}
\left[ i\hbar \gamma^{\mu} \frac{\partial}{\partial x^{\mu}} -
mc\right] S_F(x-y) = \hbar \delta^4 (x-y).
 \label{equ55}
 \end{equation}

 Now we get the modified fermionic propagator of the generalized Dirac equation. The generalized inhomogeneous Dirac equation proportional to Dirac delta function is

\begin{equation}
\left[ i\hbar \gamma^{\mu} \left( 1+\beta \hbar^2 \partial^2
\right)\frac{\partial}{\partial x^{\mu}} - mc\right] S^M_F(x) =
\hbar \delta^4 (x),
 \label{equ56}
 \end{equation}
where $S^M_F(x)$ is called the modified fermion propagator and
expressed as
\begin{equation}
 S^M_F(x) = \frac{1}{(2\pi\hbar)^4} \int d^4 p \exp\left( -\frac{i}{\hbar} p\cdot x \right)\widetilde{S}^M (p),
 \label{equ57}
 \end{equation}
where $\widetilde{S}^M (p)$ is the momentum representation of the modified fermion propagator as follows
\begin{equation}
 \widetilde{S}^M (p) = \hbar \frac{(1-\beta p^2)p + mc}{(1-\beta p^2)p^2 - m^2 c^2}.
 \label{equ58}
 \end{equation}

We note that the calculations are kept up to the first order of the deformation parameter and when $\beta$ vanishes, the modified propagator provides us with the conventional one  \cite{man, ryd}

\begin{equation}
 \widetilde{S}^M (p)|_{\beta=0} = \widetilde{S} (p).
 \label{equ59}
 \end{equation}

Using Eq. (\ref{equ58}), we have that

\begin{equation}
 S^M_F (x) = \frac{\hbar}{(2\pi \hbar)^4} \int d^4 p\exp\left(
 -\frac{i}{\hbar} p\cdot x
 \right) \frac{(1+\beta p^2)p + m c}{(1-2\beta p^2)p^2 - m^2 c^2}.
 \label{equ60}
 \end{equation}

The modified inhomogeneous Dirac equation
\begin{equation}
\left[ i\hbar \gamma^{\mu} \left( 1+\beta \hbar^2 \partial^2
\right)\frac{\partial}{\partial x^{\mu}} - mc\right] \Psi (x) =
\hbar J (x),
 \label{equ61}
 \end{equation}
has the general answer

\begin{equation}
\Psi (x) = \Psi_0 (x) + \int d^4 y S_F^M (x-y) J(y),
 \label{equ62}
 \end{equation}
where  $\Psi_0(x)$ is the answer to the modified homogeneous Dirac equation

\begin{equation}
\left[ i\hbar \gamma^{\mu} \left( 1+\beta p^2
\right)\frac{\partial}{\partial x^{\mu}} - mc\right] \Psi_0 (x) = 0,
 \label{equ63}
\end{equation}
 and the modified fermion propagator

\begin{equation}
S^M_F (x-y) = \frac{\hbar}{(2\pi \hbar)^4} \int d^4 p \exp \left(
-\frac{i}{\hbar}p\cdot (x-y) \right) \frac{1}{(1-\beta p^2)p - mc},
 \label{equ64}
\end{equation}
satisfies the following expression
\begin{equation}
\left[ i\hbar \gamma^{\mu} \left( 1+\beta \hbar^2 \partial^2
\right)\frac{\partial}{\partial x^{\mu}} - mc\right] S^M_F (x-y) =
\hbar \delta^4 (x-y).
 \label{equ65}
\end{equation}

 The modified fermion propagator $S^M_F (x - y)$  in Eq. (\ref{equ64}) can be an introduction to a modified quantum field theory to calculate the elements of matrix dispersion such as the interaction of pion with nucleon, namely the existence of a minimal observable length changes propagators and consequently the underlying field theories in high energy limit. In fact, from the viewpoint of Loop Quantum Gravity, a minimal length Generalized Uncertainty Principle leads us to a Modified Dispersion Relation (MDR) and it is the source of underlying field theory modification. In this respect, it is natural that this modification affects the interaction of pion with nucleon. In fact, the existence of a minimal length provides a natural UV cutoff for regularization of the high energy sector of the field theory (see \cite{bhat, hoss2, hoss3, hoss4, kempf, lubo, noz2, noz6, reu, stet})

 \section{Conclusion}

 The existence of the minimal observable length breaks the notion of locality and makes the space-time to be fuzzy with a foam-like structure. A fuzzy space-time gives the particle a smeared picture contrary to point-like structure of the standard model. With a smeared particle, we face modified dispersion relations. In these relations there are higher order momentum terms in the energy-momentum relation. There are severe constraints on the functional form of these modified dispersion relations coming for instance from TeV black hole thermodynamics \cite{noz4}.
 So it is natural to have modified propagators independent of the model ingredients. That is, since propagator of the underlying field theory should be modified to have regularized UV (ultra-violet) sector, for a pion-nucleon system we encounter such a modified propagator, too. These propagators usually contains both UV and IR (infra-red) sectors of the corresponding field theory safely. By using the idea of minimal length and resulting modified dispersion relation, here we introduced a generalization of the Dirac equation for free particle by the existence of a minimal length. We have considered a free fermion in a $(3+1)$-dimensional quantized space-time describing by Quesne-Tkachuk Lorentz covariant deformed algebra with $\beta , \beta^{\prime}$ as the deformation parameters. In the particular case of $\beta^{\prime} = 2\beta$ and up to the first order of parameter $\beta$, we have shown that at the Planck scale, which noncommutativity effect is a dominant property of space-time manifold at this scale, the Dirac equation includes higher order space-time derivatives of wave function. Our calculations up to the first order of $\beta$ provided us with two sets of negative and positive solutions for generalized Dirac equation. To have physically acceptable solutions, the deformation parameter $\beta$ is forced to satisfy the condition $\beta < 1/8m^2 c^2$  which is commensurate with Quesne-Tkachuk's result for a Dirac oscillator. The generalized Dirac equation added some corrections to ordinary fermionic propagator. We hope these corrections could control the divergences of the fermion quantum field theory. We  note that generally these corrections to the Dirac equation and fermionic propagator have very small order of magnitude. Nevertheless, possible realization of these effects in future highly accurate experiments provides a direct probe to test quantum gravity proposal. As we have pointed out, the modified Hamiltonian discussed in this paper usually contains momentum polynomials as the corrected terms which in the quantum domain results in the generalized Dirac equation. So, the resulting differential equation has completely different differential structure with respect to the ordinary form of the Dirac equation. This makes the problem more complex especially in the presence of the higher order momentum terms. When the order of the generalized Dirac equation increases, we encounter many mathematically possible solutions. However, imposing the physical boundary conditions, reduces the number of the acceptable solutions.






\begin{thebibliography}{999}

\bibitem{acv} D.~Amati, M.~Ciafaloni and G.~Veneziano:  \emph{Phys. Lett.} B  {\href{http://www.sciencedirect.com/science/article/pii/037026938991366X}{{\bf 216}}} (1989), 41.

\bibitem{bu}  C.~Bambi and F.~R.~Urban: \emph{Class. Quant. Grav.}  {\href{http://iopscience.iop.org/0264-9381/25/9/095006/?rss=2.0}{\bf 25}} (2008),  095006. (arXiv: {\href{http://arxiv.org/abs/0709.1965}{0709.1965}}).

\bibitem{bra1} F.~Brau:  \emph{J. Phys.} A   {\href{http://iopscience.iop.org/0305-4470/32/44/308}{\bf 32}} (1999), 7691. (arXiv: {\href{http://arxiv.org/abs/quant-ph/9905033}{quant-ph/9905033}}).

\bibitem{dv1}  S.~Das and E.~C.~Vagenas: \emph{Phys. Rev. Lett.} {\href{http://prl.aps.org/abstract/PRL/v101/i22/e221301}{\bf 101}} (2008), 221301. (arXiv: {\href{http://arxiv.org/abs/0810.5333}{0810.5333}}).

\bibitem{gar} L.~J.~Garay:  \emph{Int. J.  Mod. Phys.} A  {\href{http://www.worldscinet.com/ijmpa/10/1002/S0217751X95000085.html}{\bf 10}} (1995), 145. (arXiv: {\href{http://arxiv.org/abs/gr-qc/9403008}{gr-qc/9403008}}).

\bibitem{hoss1} S.~Hossenfelder \etal: \emph{Phys. Lett.} B  {\href{http://www.sciencedirect.com/science/article/pii/S0370269303014217}{\bf 575}} (2003), 85. (arXiv: {\href{http://arxiv.org/abs/hep-th/0305262}{hep-th/0305262}}).

\bibitem{kem1}  A.~Kempf, G.~Mangano and R.~B.~Mann: \emph{Phys. Rev.} D  {\href{http://prd.aps.org/abstract/PRD/v52/i2/p1108_1}{\bf 52}} (1995), 1108. (arXiv: {\href{http://arxiv.org/abs/hep-th/9412167}{hep-th/9412167}}).

\bibitem{kem2}  A.~Kempf: \emph{J. Phys.} A  {\href{http://iopscience.iop.org/0305-4470/30/6/030/}{\bf 30}} (1997), 2093. (arXiv: {\href{http://arxiv.org/abs/hep-th/9604045}{hep-th/9604045}}).

\bibitem{mag1}  M.~Maggiore: \emph{Phys. Lett.} B  {\href{http://www.sciencedirect.com/science/article/pii/0370269393914018}{\bf 304}} (1993), 65. (arXiv: {\href{http://arxiv.org/abs/hep-th/9301067}{hep-th/9301067}}).

\bibitem{mag2} M.~Maggiore: \emph{Phys. Lett.} B  {\href{http://www.sciencedirect.com/science/article/pii/037026939390785G}{\bf 319}} (1993),  83. (arXiv: {\href{http://arxiv.org/abs/hep-th/9309034}{hep-th/9309034}}).

\bibitem{mag3}  M.~Maggiore: \emph{Phys. Rev.} D  {\href{http://prd.aps.org/abstract/PRD/v49/i10/p5182_1}{\bf 49}} (1994), 5182. (arXiv: {\href{http://arxiv.org/abs/hep-th/9305163}{hep-th/9305163}}).

\bibitem{noz1} K.~Nozari: \emph{Phys. Lett.} B {\href{http://www.sciencedirect.com/science/article/pii/S0370269305013948}{\bf 629}} (2005), 41.
    (arXiv: {\href{http://arxiv.org/abs/hep-th/0508078}{hep-th/0508078}}).

\bibitem{noz3} K.~Nozari and T.~Azizi: \emph{Gen. Rel. Grav.}  {\href{http://www.springerlink.com/content/xgh82788m3h6585m/}{\bf 38}} (2006), 735.

\bibitem{noz7} K.~Nozari and B.~Fazlpour: \emph{Chaos, Solitons \& Fractals} {\href{http://www.sciencedirect.com/science/article/pii/S0960077906002700}{\bf 34}} (2007), 224.

\bibitem{ped}  P.~Pedram: \emph{Int. J. Mod. Phys.} D  {\href{http://www.worldscinet.com/ijmpd/19/1912/S0218271810018153.html}{\bf 19}} (2010), 2003.
    (arXiv: {\href{http://arxiv.org/abs/1103.3805}{1103.3805}}).

\bibitem{sca}  F.~Scardigli: \emph{Phys. Lett.} B  {\href{http://www.sciencedirect.com/science/article/pii/S0370269399001677}{\bf 452}} (1999), 39. (arXiv: {\href{http://arxiv.org/abs/hep-th/9904025}{hep-th/9904025}}).

\bibitem{dv2} S.~Das and E.~C.~Vagenas: \emph{Can. J. Phys.} {\href{http://www.nrcresearchpress.com/doi/abs/10.1139/P08-105}{\bf 87}} (2009), 233. (arXiv: {\href{http://arxiv.org/abs/0901.1768}{0901.1768}}).

\bibitem{kem3} A.~Kempf: On the structure of space-time at the Planck scale (1998). (arXiv: {\href{http://arxiv.org/abs/hep-th/9810215v1}{hep-th/9810215}}).

\bibitem{ben} S.~Benczik \etal: \emph{Phys. Rev.} A  {\href{http://pra.aps.org/abstract/PRA/v72/i1/e012104}{\bf 72}} (2005), 012104. (arXiv: {\href{http://arxiv.org/abs/hep-th/0502222}{hep-th/0502222}}).

\bibitem{bra2} F.~Brau and F.~Buisseret: \emph{Phys. Rev.} D  {\href{http://prd.aps.org/abstract/PRD/v74/i3/e036002}{\bf 74}} (2006), 036002. (arXiv: {\href{http://arxiv.org/abs/hep-th/0605183}{hep-th/0605183}}).

\bibitem{lub} M.~Lubo:  \emph{Phys. Rev.} D  {\href{http://prd.aps.org/abstract/PRD/v68/i12/e125004}{\bf 68}} (2003), 125004. (arXiv: {\href{http://arxiv.org/abs/hep-th/0305216}{hep-th/0305216}}).

\bibitem{que3}  C.~Quesne and V.~M.~Tkachuk:  \emph{SIGMA}  {\href{http://www.emis.de/journals/SIGMA/2007/016/}{\bf 3}} (2007), 016.
 (arXiv: {\href{http://arxiv.org/abs/quant-ph/0603077}{quant-ph/0603077}}).

\bibitem{ste}  M.~M.~Stetsko and V.~M.~Tkachuk: \emph{Phys. Rev.} A  {\href{http://pra.aps.org/abstract/PRA/v74/i1/e012101}{\bf 74}} (2006), 012101.
  (arXiv: {\href{http://arxiv.org/abs/quant-ph/0603042}{quant-ph/0603042}}).

\bibitem{noz5}  K.~Nozari: \emph{Chaos, Solitons \& Fractals} {\href{http://www.sciencedirect.com/science/article/pii/S0960077906006485}{\bf 32}} (2007), 302.

\bibitem{que1} C.~Quesne and V.~M.~Tkachuk: \emph{J. Phys.} A  {\href{http://iopscience.iop.org/0305-4470/39/34/021/}{\bf 39}} (2006), 10909.
 (arXiv: {\href{http://arxiv.org/abs/quant-ph/0604118}{quant-ph/0604118}}).

\bibitem{que2} C.~Quesne and V.~M.~Tkachuk: \emph{Czech. J. Phys.} {\href{http://www.springerlink.com/content/1502h17556121nln/}{\bf 56}} (2006), 1269.
 (arXiv: {\href{http://arxiv.org/abs/quant-ph/0612093}{quant-ph/0612093}}).

\bibitem{sny}  H.~S.~Snyder: \emph{Phys. Rev.} {\href{http://prola.aps.org/abstract/PR/v71/i1/p38_1}{\bf 71}} (1947), 38.

\bibitem{gre}  A.~E.~S.~Green: \emph{Phys. Rev.} {\href{http://prola.aps.org/abstract/PR/v73/i1/p26_1}{\bf 73}} (1948), 26.

\bibitem{grein}  W.~Greiner: \emph{Relativistic Quantum Mechanics: Wave Equations}, 3rd ed., Springer, Berlin, 2000.

\bibitem{man}  F.~Mandl and G.~Shaw: \emph{Quantum Field Theory}, Wiley, New York 1984.

\bibitem{ryd}  L.~H.~Ryder: \emph{Quantum Field Theory}, 2nd ed., Cambridge University Press, Cambridge 1996.

\bibitem{bhat}  G.~Bhattacharyya \etal: \emph{Phys. Lett.} B  {\href{http://www.sciencedirect.com/science/article/pii/S0370269304014364}{\bf 603}} (2004), 46.
    (arXiv: {\href{http://arxiv.org/abs/hep-ph/0408295}{hep-ph/0408295}}).

\bibitem{hoss2} S.~Hossenfelder: \emph{Phys. Rev.} D  {\href{http://prd.aps.org/abstract/PRD/v70/i10/e105003}{\bf 70}} (2004), 105003.
   (arXiv: {\href{http://arxiv.org/abs/hep-ph/0405127}{hep-ph/0405127}}).

\bibitem{hoss3}  S.~Hossenfelder: \emph{Phys. Rev.} D  {\href{http://prd.aps.org/abstract/PRD/v73/i10/e105013}{\bf 73}} (2006), 105013.
 (arXiv: {\href{http://arxiv.org/abs/hep-th/0603032}{hep-th/0603032}}).

\bibitem{hoss4}  S.~Hossenfelder: \emph{Class. Quant. Grav.}  {\href{http://iopscience.iop.org/0264-9381/25/3/038003/}{\bf 25}} (2008), 038003.
 (arXiv: {\href{http://arxiv.org/abs/0712.2811}{0712.2811}}).

\bibitem{kempf} A.~Kempf and G.~Mangano: \emph{Phys. Rev.} D  {\href{http://prd.aps.org/abstract/PRD/v55/i12/p7909_1}{\bf 55}} (1997), 7909.
 (arXiv: {\href{http://arxiv.org/abs/hep-th/9612084}{hep-th/9612084}}).

\bibitem{lubo} M.~Lubo: \emph{Phys. Rev.} D {\href{http://prd.aps.org/abstract/PRD/v61/i12/e124009}{\bf 61}} (2000), 124009.
 (arXiv: {\href{http://arxiv.org/abs/hep-th/9911191}{hep-th/9911191}}).

\bibitem{noz2} K.~Nozari and M.~Karami: \emph{Mod. Phys. Lett.} A  {\href{http://www.worldscinet.com/mpla/20/2040/S0217732305018517.html}{\bf 20}} (2005), 3095.
 (arXiv: {\href{http://arxiv.org/abs/hep-th/0507028}{hep-th/0507028}}).

\bibitem{noz6}  K.~Nozari and S.~H.~Mehdipour: \emph{Chaos, Solitons \& Fractals}  {\href{http://www.sciencedirect.com/science/article/pii/S0960077906008824}{\bf 32}} (2007), 1637.
    (arXiv: {\href{http://arxiv.org/abs/hep--th/0601096}{hep-th/0601096}}).

\bibitem{reu}   M.~Reuter and J.~Schwindt: \emph{J. High Energy Phys.} {\href{http://iopscience.iop.org/1126-6708/2006/01/070/}{\bf 01}} (2006), 070.
  (arXiv: {\href{http://arxiv.org/abs/hep-th/0511021}{hep-th/0511021}}).

\bibitem{stet} M.~M.~Stetsko and V.~M.~Tkachuk: \emph{Phys. Rev.} A {\href{http://pra.aps.org/abstract/PRA/v76/i1/e012707}{\bf 76}} (2007), 012707.
 (arXiv: {\href{http://arxiv.org/abs/hep-th/0703263}{hep-th/0703263}}).

\bibitem{noz4}  K.~Nozari and A.~S.~Sefidgar: \emph{Phys. Lett.} B  {\href{http://www.sciencedirect.com/science/article/pii/S0370269306002310}{\bf 635}} (2006), 156.
 (arXiv: {\href{http://arxiv.org/abs/gr-qc/0601116}{gr-qc/0601116}}).

\end{thebibliography}
\end{document}